\newif\ifabstract
\abstracttrue   
\abstractfalse   

\ifabstract  

\documentstyle[twocolumn]{article}
\makeatletter
\def\section{\@startsection {section}{1}{\z@}{-3.5ex plus -1ex minus
      -.2ex}{2.3ex plus .2ex}{\large\bf}}
\makeatother

\else  

\documentstyle[12pt]{article}

\fi 

\def\8{\infty}
\def\oh{\frac{1}{2}}

\def\i{\imath\,}

\def\undertext#1{\vtop{\hbox{#1}\kern 1pt \hrule}}

\def\lrb#1{\left(#1\right)}

\def\VEV#1{\left\langle\,#1\,\right\rangle}

\def\dbyd#1#2{\frac{d#1}{d#2}}
\def\pp#1{\frac{\partial}{\partial#1}}
\def\pbyp#1#2{\frac{\partial#1}{\partial#2}}

\def\fbyf#1#2{\frac{\delta#1}{\delta#2}}

\def\br{\\ \nonumber & &}

\def\be{\begin{equation}}
\def\ee{\end{equation}}
\def\bea{\begin{eqnarray} & &}
\def\eea{\end{eqnarray}}
\def\ct#1{\cite{#1}}
\def\rf#1{(\ref{#1})}
\def\EXP#1{\exp\left(#1\right)}

\def \PDF{probability distribution function }
\def\t{\tilde}
\author{
{\sc Victor Gurarie}\\
  {\footnotesize and}\\[-0.7ex]
{\sc Alexander Migdal}\\
  {\footnotesize Department of Physics}\\[-0.7ex]
  {\footnotesize Princeton University}\\[-0.7ex]
  {\footnotesize Princeton, NJ 08544}
}

\title{Instantons in Burgers Equation}

\begin{document}
 \maketitle
\begin{abstract}

The instanton solution for the forced Burgers equation is found. This
solution describes the exponential tail of the probability
distribution function of velocity
differences in the region where shock waves are absent. The
results agree with the one found recently by Polyakov, who used the
operator product conjecture. If this conjecture is true,
then our WKB asymptotics of the Wyld functional integral is
exact to all orders of perturbation expansion around the instanton
solution. We explicitly checked this in the first order.
We also generalized our solution for the arbitrary dimension  of
Burgers (=KPZ) equation. As a result we found the angular dependence
of the velocity difference PDF.

\end{abstract}

There are two complementary views of the turbulence problem. One could
regard it as kinetics, in which case the time dependence of velocity
\PDF (PDF) must be studied. The Wyld functional integral describes the
correlations functions of the velocity field in this picture.

Another view is the Hopf (or Fokker-Planck) approach, where the equal
time PDF is studied. For the random force distributed as white noise
in time, the closed functional equations (Fokker-Planck equation) can
be derived. In the case of thermal noise the Boltzmann distribution
can be derived as an asymptotic solution of this equation.

One of the authors \ct{Loop} reduced the Hopf equations for the full
Navier-Stokes equation to the one dimensional functional equation
(loop equation). The WKB solutions of this equation were studied,
leading to the area law for the velocity circulation PDF.

In the recent paper by Polyakov
\ct{Pol} a similar method of solving the randomly
driven Burgers equation was proposed.  It reduced the problem of
computations of their correlation functions to the solution of a
certain partial differential equation. This equation for the velocity
difference PDF can be explicitly solved.

The derivation of the Polyakov equation was based on the conjecture of the
existence of the operator product expansion.

On the other hand, it was recently understood in the
paper \ct{FLM} by Falkovich, Kolokolov, Lebedev and one of us
 that the Wyld
functional integral provides the general method of computation of the
PDF tails. One has to find an {\em instanton} -- the minimum of the
action in the Wyld functional integral.

This instanton is not the same as the solution of the classical
equation in the usual sense. The force is present, and it acts
in a self-consistent way, as required by minimization of the
action. This force is no longer random, it is adjusted to provide the
large fluctuation of velocity field under consideration.

It was shown in \ct{FLM} that the PDF for the passive scalar advection in the
Gaussian velocity field is asymptotically described by an instanton,
with the spatially homogeneous strain.

Here we find instanton in the Burgers equation, in the presence of finite
viscosity. The result which we obtain in the turbulent limit
(vanishing viscosity) coincides with that of Polyakov which gives
an indirect confirmation to his OPE conjecture.

We start with the randomly driven Burgers equation.
\be
\label{EQburg}
u_t+u u_x -\nu u_{xx} = f(x,t)
\ee
where the force $f(x,t)$  is a gaussian random field with a
pair correlation function
\be
\VEV{f(x,t) f(y,t')}  = \delta(t-t') \kappa(x-y)
\ee

The Wyld functional integral has the following form (see, for example,
\ct{Jus}).
\bea
\label{EQjustin}
\int {\cal D} f \EXP{ -\oh \int dx dy dt \ f(x,t) D(x-y) f(y,t)}  = \br
\int {\cal D} f {\cal D} u \   \delta [u_t+u u_x -\nu u_{xx}-f]
\EXP{ -\oh \int dx dy dt \ f(x,t) D(x-y) f(y,t)} = \br
\int {\cal D} f {\cal D} u {\cal D} \mu   \EXP{ \i \int d x d t \mu (
u_t+u u_x -\nu u_{xx}-f) - \oh \int d x d y d t f(x,t) D(x-y) f(y,t)}  =  \br
\int {\cal D} u {\cal D} \mu \ \EXP{ \i \int d x d t \mu (
u_t+u u_x -\nu u_{xx})- \oh \int dx dy dt  \ \mu(x,t) \kappa(x-y) \mu(
y,t)}
\eea
Here we started with the obvious functional integral for the force, where $D$
is the function inverse to $\kappa$. To change the variable of
integration from $f$ to
 $u$ we inserted into the functional integral the identity
\be
\label{EQid}
 {\cal N} =\int{\cal D} u \ \delta \left[ u_t+u u_x -\nu u_{xx}-f
\right]
\ee
where ${\cal N}$ is just a number. Dropping that number as an
unimportant constant,  we removed the delta function at the expense
of introducing a ``conjugated'' variable $\mu$ and evaluated
the integral over $f$ to arrive at the final expression in \rf{EQjustin}.

It is not so obvious that the integral in \rf{EQid} is equal to  a pure
number because of the determinant $\det \lrb{\fbyf{f}{u}}$ arising in its
computation. It is possible, however, to prove \rf{EQid} using the
causality argument (see  \ct{Jus}).

Thus the initial problem of computing the correlation functions of
Burgers equation is reduced to the field theory with the action
\be
\label{EQactionold}
 S= - \i \int \mu (
u_t+u u_x -\nu u_{xx})+ {1 \over 2} \int dx dy dt  \ \mu(x,t)
\kappa(x-y) \mu(
y,t)
\ee

Here we are going to study the correlation function
\be
\label{EQcorr}
\VEV{ \exp\left\{ \lambda_0 (u(\rho_0/2)-u(-\rho_0/2))
\right\} } = \int {\cal D}u {\cal D} \mu \exp\left\{ \lambda_0
(u(\rho_0/2)-u(-\rho_0/2)) - S\right\}
\ee
whose Laplace Transform gives us the two point probability distribution
(see \ct{Pol}). We will often refer to the expression we have
in the exponential as the action
\be
\label{EQaction}
 S_{\lambda_0} = S - \lambda_0
(u(\rho_0/2)-u(-\rho_0/2))
\ee
There are no general methods to compute the functional integrals like
\rf{EQcorr}
exactly. The most straightforward approach would be to expand the
exponential in  functional
integral in powers of the nonlinear term $\mu u u_x$. By doing so
we will just reproduce the well known Wyld's
diagram technique (see \ct{Wyl}).  The attempts to use this technique
to describe turbulence always failed because we are interested in the
limit $\nu \rightarrow 0$ when nonlinear term dominates the  functional
integral.
The absence of a large parameter makes the task of computing this functional
integral by using perturbation theory hopeless.

Nevertheless, if we are interested in computing the large $\lambda_0$
behavior of the correlation function in \rf{EQcorr}, we can
use $\lambda_0$ itself as a large parameter. Then the integral will
be dominated by its saddle point, or by the solutions of the
equations of motion for the
action \rf{EQaction}. All we have to do is to find those solutions and
compute the value of the action $S_{\rm inst}$
on those solutions. The answer will be given by
\be
\label{EQanswer}
\VEV { \exp\left\{ \lambda_0 (u(\rho_0/2)-u(-\rho_0/2))
\right\} }= \exp \left\{ -S_{\rm inst} \left( \lambda_0 \right)
+S_{\rm inst} \left( 0 \right) \right\}.
\ee
If we want, we can then further expand the integral in powers of
$1/\lambda_0$ using the perturbation theory around those
solutions. We will call this method WKB approximation and the solutions
instantons using the names borrowed from quantum field theory.

To that effect, let us write down the equations of motion corresponding
to the action \rf{EQaction}. They are
\be\label{EQmotiono}
 u_t + u u_x - \nu u_{xx}  = - \i \int dy  \
\kappa (x-y)
\mu(y)
\ee
\be
\label{EQmotiont}
\mu_t + u \mu_x + \nu \mu_{xx}  = - \i \lambda_0 \left\{
\delta \left( x-{\rho_0 \over 2} \right) -
\delta \left( x+{\rho_0  \over 2} \right) \right\} \delta (t)
\ee

To solve these equations, let us first notice that the only role the
right hand side of \rf{EQmotiont} plays is
giving the field $\mu$ a finite discontinuity at $t=0$. It is also easy to
see that $\mu(t)=0$ for $t>0$.  This is because $\mu$ feels a negative
viscosity
so any solution which is nonzero at $t>0$ will become singular\footnote{
Alternatively, one can argue that the integrals in \rf{EQjustin} are
defined only for $t<0$. Those arguments use a striking similarity
between \rf{EQjustin} and a Feynman path integral for a quantum
mechanical system with the coordinates $u$ and momenta $\mu$ to
define \rf{EQcorr} as a wave function in the momentum representation.
Then the conditions \rf{EQmu} become obvious.}.
Thus the field $\mu$ can be
evaluated
at $t=-0$ to be
\be
\label{EQmu}
\mu(t=-0)= \i \lambda_0 \left\{ \delta \left( x-{\rho_0 \over 2}
\right) -
\delta \left( x+{\rho_0  \over 2} \right) \right\}
\ee
while it is zero at all later moments of time. It is therefore convenient
 to speak of the field $\mu$ propagating backwards in time
starting from its initial value given by \rf{EQmu}.

If we try to propagate \rf{EQmu} back in time, we discover that
we have to deal with two phenomena governed by the second and third
terms of \rf{EQmotiont}. One of them is just
a motion of the initial conditions as dictated by the velocity $u$. The
other is the ``smearing" of the initial delta function distributions in
\rf{EQmu} due to the viscosity.

However, it can be shown by a direct computation that the smearing does
not change the value of the action on the instanton as long as the
viscosity
is not very large. We will construct a solution which takes into
account the viscosity in the end of the paper.
For now
we will just drop the viscosity term to arrive
at a simplified equation
\be
\label {EQmotionmu}
\mu_t+ u \mu_{xx} =0
\ee
Since all this equation can do is moving the $\delta$ function-like
singularities around (and changing their heights  by compressing them),
it is clear that the solution of \rf{EQmotionmu} with the
boundary conditions given by
\rf{EQmu} is just
\be
\label{EQmut}
 \mu(t)= \i \lambda(t) \left\{ \delta \left( x-{\rho(t) \over 2} \right) -
\delta \left( x+{\rho(t)  \over 2} \right) \right\}
\ee
with the boundary conditions
\be
\label{EQbound}
 \lambda(0)=\lambda_0, \ \ \rho(0)=\rho_0
\ee

Now let us leave the equation \rf{EQmotiont} for a while and
study \rf{EQmotiono}. A natural thing to do is to substitute \rf{EQmut}
into the right hand side of \rf{EQmotiono}.

We obtain
\be
\label{EQmotu}
 u_t+ u u_x - \nu u_{xx} = \lambda \left\{
\kappa \left( x-{\rho \over 2} \right)-\kappa \left( x+{\rho \over 2}
\right) \right\}
\ee

To proceed further we need to know $\kappa$. Let us assume, following
\ct{Pol}, that $\kappa(x)$ is a slow varying even function of $x$
which behaves as
\be
\label{EQforkappa}
\kappa(x)\approx \kappa(0)-{\kappa_0 \over 2} x^2, \ \
|x| \ll
\sqrt { \kappa(0) \over \kappa_0 } \equiv L
\ee
and quickly turns into zero when $|x| \gg L$. The interval $L$
characterizes the range of  the random force and
we will work only there, that is we suppose that $\rho$ also lies
within this interval. It is clear then that the contribution to the
action \rf{EQaction} comes only from the interval $L$  (compare
with \rf{EQmut}). So we do not have to know the velocity beyond that
interval. There we use \rf{EQforkappa} to obtain
\be
\label{EQeqforu}
 u_t + u u_x - \nu u_{xx} = \lambda  \rho \kappa_0 x
\ee
Notice that $\kappa(0)$ dropped out.

\rf{EQeqforu} is a Burgers equation with a linear force. It is easy to
solve such an equation. We have to look for the solution in terms of
a linear function
\be\label{EQlin}
u(x,t)=\sigma(t) x
\ee
which leads to
\be
\label{EQforsigma}
 {d \sigma \over d t} + \sigma^2 =
\kappa_0 \lambda \rho
\ee
Notice that the viscosity term did not contribute. That does not mean
that the viscosity is not important at all. For $x \gg \rho$  the
force in the \rf{EQmotu} becomes zero and the viscosity  there is
important to make the velocity go to zero at the infinity. However,
in the region $x  \propto \rho$  which is the one we study the viscosity
term can be dropped.

Now we can use \rf{EQlin} and \rf{EQmut}
to solve  \rf{EQmotionmu}. A direct
substitution leads to
\bea
\label{EQmotionall}
{d \lambda \over d t}  = \lambda \sigma \br
           {d \rho    \over d t}  = \rho    \sigma
\eea
These can be solved in terms of the function
$$R(t)=\exp\left( \int_0^t dt^\prime \sigma(t^\prime)\right) $$
to give
\bea
\label{EQboth}
 \lambda  = \lambda_0 R \br
 \rho     = \rho_0    R
\eea
while $R$ itself satisfies, by virtue of \rf{EQforsigma}, the equation
\be
\label{EQforar}
 {d^2 R \over dt^2} = \kappa_0 \rho_0 \lambda_0 R^3
\ee
The last equation has to be solved with the boundary condition
$R(-\infty)=0$ otherwise the action \rf{EQaction} will not be finite. The
solution is given by
\be
\label{EQrbeha}
  R={1 \over 1 - \sqrt { \kappa_0 \rho_0 \lambda_0 \over 2} t}
\ee
So we have found the instanton solution for the equations \rf{EQmotiono},
\rf{EQmotiont}. Notice that it is the {\sl only} solution of the
equations of motion, so we do not have to sum over different
instantons.

Now it is a matter of a simple computation to find the action on the
instanton.
We collect everything together and substitute \rf{EQrbeha},
\rf{EQboth}, \rf{EQmut}, and \rf{EQlin} back to \rf{EQaction} to get
\be
\label{EQinstanton}
S_{\rm inst}= - { \sqrt {2 \kappa_0} \over 3}
(\lambda_0 \rho_0)^{3 \over 2}
\ee
while the correlation function we were studying is
\be
\label{EQotvet}
 \VEV { \exp\left\{ \lambda_0 \left( u(\rho_0/2)-u(-\rho_0/2) \right)
\right\} } = \exp \left( { \sqrt {2 \kappa_0} \over 3}
\left(\lambda_0 \rho_0 \right)^{3 \over 2} \right)
\ee
This is the same answer as the one obtained in \ct{Pol}. We want to
emphasize, however, that we obtained it without any conjectures and
only as an asymptotics for $|\lambda_0\ | \gg 1$.

We are not going to  discuss the physical implications of \rf{EQotvet}
referring instead to the papers \ct {Pol} and \ct {Yakh}.

Now we return to the question of why we can drop the viscosity in
\rf{EQmotiont}. Namely, we just construct the solution of \rf{EQmotiont}
with the viscosity. To do that, it is convenient to Fourier
transform it
(taking into account that the velocity $u$ is a linear
function of $x$).
\be
\label{EQmuvisc}
\pbyp{\mu(p)}{t} - \sigma \pp{p} \left( p \mu(p) \right) - \nu \ p^2 \mu =0
\ee

Then the solution of \rf{EQmuvisc} can be found as a direct
generalization of \rf{EQmut}.
\be
\label{EQmuviscsol}
\mu(p) \propto \lambda(t) \sin \left( {p \ \rho(t) \over 2} \right) \exp
\left( - \beta(t) p^2 \right)
\ee
Here we had to introduce the new variable $\beta(t)$ which
measures the speed of smearing of the solution with the evident
initial condition $\beta(0)=0$. Substituting \rf{EQmuviscsol} into
\rf{EQmuvisc} we reproduce the equations \rf{EQboth} for $\rho$ and
$\lambda$ with the additional equation for $\beta$
\be
\beta_t - 2 \sigma  \beta + \nu =0
\ee
with the solution
\be
\label{EQbeta}
\beta(t)= {\nu \over 3 \omega} { (1- \omega t)^3 -1 \over
{\left( 1- \omega t \right) }^2 }
\ee
where $\omega=\sqrt {(\kappa_0 \rho_0 \lambda_0) /2}$.

The smearing, however, has no influence whatsoever on the velocity.
To see that, we substitute \rf{EQmuviscsol}
into \rf{EQmotiono} and arrive back
at \rf{EQeqforu}. In other words, the variable $\sigma$ still satisfies
the
same equation \rf {EQforsigma}.
A simple argument given below shows that the
value of the instanton action depends only on the final value
of the velocity of the instanton solution which, as we just showed,
does not depend on the viscosity.

We must remember, though, that we should not allow
$\mu$ to spread beyond $L$ interval, or more precisely
$L> \sqrt \beta$. $\beta$ can become arbitrarily large for large
negative times, but
the characteristic time interval which contributed to
the computation of the instanton action is
\be
 t_{\rm inst}={1 \over \omega}
\ee
So all we have to do is to make sure that $ \beta(t_{\rm inst}) < L^2$
or
\be
L^2 > { \nu \over \omega}
\ee
This is the condition which viscosity must satisfy for \rf{EQotvet}
to be correct and independent of viscosity.

So far we cannot claim that \rf{EQotvet} is an exact answer. It is
just a leading asymptotic if $|\lambda_0|$ is a large number. It might
be important to estimate the next order contribution to the \rf{EQotvet}
especially in view of the claim made in \ct{Pol} that \rf{EQotvet} is
actually exact.

To do that it is convenient to introduce the quantity
\be
\label{EQderiv}
\pp{\lambda_0} \log
\VEV { \exp\left\{ \lambda_0 (u(\rho_0/2)-u(-\rho_0/2))
\right\} } = { \int {\cal D} u {\cal D} \mu
\left( u(\rho_0/2)-u(-\rho_0/2)  \right) \exp
\left( -S_{\lambda_0} \right) \over
\int {\cal D} u {\cal D} \mu
\exp
\left( -S_{\lambda_0} \right) }
\ee
It is easy to expand this quantity around the instanton solution.
Writing $u = u_{\rm inst}+\tilde{u}$, $\mu = \mu_{\rm inst}
+ \t{\mu}$ we arrive for \rf{EQderiv}
at
\be
\label{EQjustla}
u_{\rm inst} \left({ \rho_0 \over  2 } \right) - u_{\rm inst}
\left( - {\rho_0 \over  2}
\right) +
{ \int {\cal D} \t{u} {\cal D} \t{\mu}
\left( \t{u}(\rho_0/2)-\t{u}(-\rho_0/2)  \right) \exp
\left\{ -S_{\lambda_0} \left( u_{\rm inst} +\t{u} ,
\mu_{\rm inst} +\t{\mu} \right) \right\} \over
\int {\cal D} \t{u} {\cal D} \t{\mu}
\exp
\left\{ -S_{\lambda_0}  \left( u_{\rm inst}+\t{u},
\mu_{\rm inst} +\t{\mu} \right) \right\} }
\ee

The first term of the expression corresponds to the instanton
contribution to the action. We see that it depends only on the
value of the instanton solution at $t=0$, justifying our argument
that the viscosity does not contribute to the answer. We can easily
compute this term by substituting the known instanton solution to get
\be
\label{EQformul}
 \sqrt{\kappa_0 \lambda_0 \over 2} \rho_0^{3 \over 2}
\ee
 which is of course compatible with  \rf{EQotvet}.

Let us proceed to estimate the second term in \rf{EQjustla}.
It is not difficult to see that if we expand
$S_{\lambda_0} \left(u_{\rm inst}+\t{u},
 \mu_{\rm inst} +\t{\mu} \right)$ in powers of $\t{u}$ and
$\t{\mu}$ up to a second order,
the contribution from
that will be zero as $S$ will be an even function of $\t{u}$ and $\t{\mu}$.
However, due to the presence of the third order terms in the
action, it is not clear if the higher order terms are also zero.
If they are nonzero, then  a dimensional argument shows that
 the next term in
\rf{EQformul} is of the order $(1/\lambda_0)^{2}$.
Really, then the expanded action looks like
\bea
\label{EQexpand}
S(u_{\rm inst}+\t{u}, \mu_{\rm inst}+\t{\mu}) =
S_{\rm inst}+S_{\rm quad}+S_{\rm
int} = S_{\rm inst} -
  \i \int \t{\mu} (
\t{u}_t -\nu \t{u}_{xx})+ \br {1 \over 2} \int dx dy dt  \ \t{\mu}(x,t)
\kappa(x-y) \t{\mu}(
y,t) - \i  \int \t{\mu} \pp{x} \left( \t{u} u_{\rm inst} \right) - \i
\int \mu_{\rm inst} \t{u} \t{u}_x  + S_{\rm int}
\eea
$S_{\rm quad}$ meaning the quadratic part of the expansion and
$S_{\rm int}$ being the interaction term,
$S_{\rm int}=- \i \int \t{\mu} \t{u} \t{u}_x$.
We need to  take into account the interaction in the first order
of perturbation theory. That is  we need to evaluate
\be
\label{EQcontt}
\left\langle \t{u} \left( \rho_0 \right)  S_{\rm int} \right\rangle
\ee
understanding the average in the sense of the action
$S_{\rm quad}$. While the ``honest'' computation would require the
computation of the Green's functions for that action, we
can  estimate the value of \rf{EQcontt} by noting that $u_{\rm inst}
\propto \sqrt {\lambda_0}$, $\mu_{\rm inst} \propto \lambda_0$ and
\be
S_{\rm quad} \propto \lambda_0 \  \t{u}^2 +
\sqrt{\lambda_0} \ \t{\mu} \t{u}
\ee
Therefore the typical fluctuations are $\delta \t{u} \propto
{1 \over \sqrt{\lambda_0}}$ and $\delta \t{\mu} \propto 1$.
The expression \rf{EQcontt} involves three $u$ and one $\mu$, so
we could expect it to be of the order $1/\lambda_0^{3 \over 2}$. However,
there is also an integration over $t$ involved. That integration
gives us $1/\omega \propto 1/\sqrt{\lambda_0}$. Altogether we
arrive at $1/\lambda_0^2$.

In other words, it could be expected that  the correction to the answer
has the form
\be
\label{EQotvetxy}
 \VEV { \exp\left\{ \lambda_0 \left( u(\rho_0/2)-u(-\rho_0/2) \right)
\right\} } \propto \exp \left( { \sqrt {2 \kappa_0} \over 3}
\left(\lambda_0 \rho_0 \right)^{3 \over 2} + {{\rm const}
\over \lambda_0}
\right)
\ee
However, this question requires much deeper investigations which we
leave for future work. The formula \rf{EQotvetxy} is just a dimensional
estimation. The Green's functions for $S_{\rm quad}$ have to be constructed
for the corrections to the answer \rf{EQotvet} to be computed reliably.
That has not been done yet.

The analysis of this paper can easily be extended for the case
of more than one dimension. The analog of the equations \rf{EQmotiono}
and \rf{EQmotiont} will be
\be\label{EQmotionod}
\pbyp{ u_i}{t}  +  \left( u_j \pp{x_j} \right) u_i - \nu \Delta u  =
- \i \int dy  \
\kappa_{i j} (x-y)
\mu_j (y)
\ee
\be
\label{EQmotiontd}
\pbyp{\mu_i}{t} + \pp{x_j} \left( u_j \mu_i \right) -
\mu_j \pbyp{u_j}{x_i}
 + \nu \Delta \mu  = - \i {\lambda_0}_i \left\{
\delta \left( x-{\rho_0 \over 2} \right) -
\delta \left( x+{\rho_0  \over 2} \right) \right\} \delta (t)
\ee

The solution  of these equations is a direct generalization of
\rf{EQmut}, or
\be
\label{EQmutd}
 \mu_i (t)= \i \lambda_i (t) \left\{ \delta \left( x-{\rho(t) \over 2} \right)
-
\delta \left( x+{\rho(t)  \over 2} \right) \right\}
\ee
with
\be
\label{EQboundd}
 \lambda_i(0)={\lambda_0}_i , \ \ \rho_i(0)={\rho_0}_i
\ee

The further progress depends on the tensorial structure of $\kappa_{ij}$
which is just a correlation function
\be
\langle f_i (x,t) f_j(y,t') \rangle = \kappa_{ij}(x-y) \delta(t-t')
\ee
A natural thing to assume would be that the force is a gradient of
something, that is $f_i=\partial_i \Phi$, in which case
\be
\kappa_{i j} (x) \approx \kappa_{ij}(0) - {\kappa_0 \over 2}
(x^2 \delta_{ij} + 2 x_i
x_j)
\ee

A direct generalization of the velocity ansatz is
\be
u_i=\sigma_{ij} x_j
\ee
and the equations  \rf{EQforsigma} and \rf{EQmotionall} turn into
\bea
\dbyd{\lambda_i}{t} = \lambda_j \sigma_{ji} \br
\dbyd{\rho_i}{t}    = \sigma_{ij} \rho_j    \br
\dbyd{\sigma_{i j}}{t} + \sigma_{i k} \sigma_{k j} = \kappa_0 \left(
\lambda_i \rho_j +
\rho_i \lambda_j +  \delta_{i j} \lambda_k \rho_k
\right)
\eea

$\sigma$ can actually be eliminated from those equations to give us an
analog of  \rf{EQforar},
\bea
\label{EQnad}
{ d^2 \lambda_i \over dt^2 } = \kappa_0 \left( \rho_i \lambda^2
+ 2 \lambda_i \lambda_k \rho_k \right) \br
{d^2 \rho_i \over dt^2 } = \kappa_0 \left( \lambda_i \rho^2
+ 2 \rho_i \lambda_k \rho_k \right)
\eea

While a general solution of those equations is rather difficult to
find, it is possible to find  the action on the solution by
analyzing the corresponding Hamilton-Jacobi equation. To do that,
we note that the equations \rf{EQnad} are hamiltonian with the
hamilton function
\be
H=\dbyd {\lambda_i}{t} \dbyd {\rho_i}{t} - {\kappa_0 \over 2}
\left( \rho^2 \lambda^2 + 2 {\left( \lambda_k \rho_k \right)}^2
\right)
\ee
The (time independent) instanton action $S$ clearly satisfies the
equation \footnote{We would like to note that the equation
\rf{EQpol}
follows
from the master equation of \ct{Pol} if the viscosity is
completely neglected. The authors are grateful to A. Polyakov
for pointing that out.}
\be
\label{EQpol}
\pbyp {S} {\lambda_i} \pbyp {S} {\rho_i} - {\kappa_0 \over 2}
\left( \rho^2 \lambda^2 + 2 {\left( \lambda_k \rho_k \right)}^2
\right) =0
\ee

By rescaling the time and the variables $\rho$ and $\lambda$
in \rf{EQnad} we can show that the
action $S$ has the following initial condition dependence
\be
S=\sqrt
{ \kappa_0 \over 2} {\left( \rho_0 \lambda_0 \right) }^{3 \over 2}
 f \left( \cos \varphi \right)
\ee
where $\varphi$ is the angle between the vectors of initial conditions
${ \lambda_0 }_i$ and
${  \rho_0  }_i$. This ansatz coincides with the
one dimensional answer \rf{EQinstanton} up to a nontrivial
function of the angle $f\left( \cos \varphi \right)$ which we would like to
determine.
Plugging the ansatz into the
Hamilton-Jacobi  equation we obtain the equation for $f(\cos \varphi)$
\be
\label{EQstrashny}
{9 \over 4} z {f}^2 + 3 f f' (1-z^2) +{f'}^2 (-z+z^3) =  1+2
z^2
\ee
where $z=\cos(\varphi)$. This first order differential equation
has to be solved with the boundary condition
\be
f(1)={2 \over \sqrt{3}}
\ee
which follows directly from the equation \rf {EQstrashny} but also can
be computed by solving the equation of motion for $\varphi=0$.
We can find the function $f$ as a series in powers of $1-z$. It turns
out there are two solutions
\bea
\label{EQforf}
f(z)=
{2 \over \sqrt{3}}-{\sqrt{3}+\sqrt{11} \over 4} (1-z)+
{5 \sqrt{33} - 61 \over 32 \left( 3 \sqrt{3} - 2 \sqrt{11} \right) }
{(1-z)}^2  +\dots \br
f(z)={2 \over \sqrt{3}}+ {\sqrt{11} - \sqrt{3} \over 4} (1-z) -
{ 5 \sqrt{33} + 61 \over 32 \left( 3 \sqrt{3} + 2 \sqrt{11} \right) }
{(1-z)}^2 + \dots
\eea

The equation \rf{EQstrashny} does not tell us which of these two
to choose. We have to match \rf{EQforf} with the solution of
\rf{EQnad}. Those equations cannot be solved in general, but there is
a way to find their solution if the angle $\varphi$ is close to zero
which should be enough to determine $f'(1)$ and therefore to choose
the right action.

To do that, we note that the motion represented by \rf{EQnad} is
essentially two dimensional, with all the motion confined to the
$ \lambda_0 ,  \rho_0$ plane. Then we represent $\lambda$
as a two-vector $(\lambda_1, \lambda_2)$, while $\rho = (\lambda_1,
-\lambda_2)$. The equations \rf{EQnad}  turn into (we choose the units
where $\kappa_0=1$)
\bea
{ d^2 \lambda_1 \over dt^2 } = 3 \lambda_1^3 - \lambda_1 \lambda_2^2
\br
  { d^2 \lambda_2 \over dt^2 } = - 3 \lambda_2^3 + \lambda_2
\lambda_1^2
\eea

We choose the boundary conditions $\lambda_0=1,  \ \rho_0=1$.
If $\varphi=0$ then $\lambda_2=0$ while $\lambda_1$ satisfies the
equation
\be
{ d^2 \lambda_1 \over dt^2 } = 3 \lambda_1^3
\ee
hence
\be
\lambda_1={1 \over 1 - \omega t}, \ \ \omega=\sqrt{3 \over 2}
\ee
Now if $\varphi$ is a small number then $\lambda_2 \ll \lambda_1$ and
it satisfies  the approximate equation
\be
{ d^2 \lambda_2 \over dt^2 } =
{\lambda_2 \over {\left(1 - \omega t\right)}^2}
\ee
with the solution
\be
\lambda_2={C \over {\left(1 - \omega t\right)}^\alpha}, \ \
\alpha={-3 \pm \sqrt{33} \over 6}
\ee
In particular, for $t=0$,
\be
\label {EQlogder1}
\dbyd{\log \lambda_2}{t}  = {-\sqrt{3} \pm \sqrt{11} \over 2 \sqrt{2}}
\ee
That last quantity can also be evaluated if we know the action $S$.
For $t=0$ we obtain
\be
\dbyd{\rho_i}{t}=\pbyp{S}{\lambda_i}={ 3 \lambda_i \over 2 \sqrt{2}}
f(z) + {f'(z) \over \sqrt{2}} \left( \rho_i - z \lambda_i
\right)
\ee
which translates to the language of $\lambda_2$ ($\lambda_2 \ll
\lambda_1$ and $z \approx 1$) as
\be
\label{EQlogder2}
\dbyd{\log \lambda_2}{t} = \sqrt{2} f'(1) - \sqrt{3 \over 2}
\ee
Comparing with \rf{EQlogder1} we get
\be
f'(1) = {\sqrt{3} \pm \sqrt{11} \over 4}
\ee

Now we need to choose the plus sign in all the above formulas
as we want $\alpha$ to be positive. Otherwise our action will correspond
to the solution growing at $t \rightarrow -\infty$. That makes us
choose the first $f(z)$ in \rf{EQforf}.

We would also like to note that according to the equation
\rf{EQstrashny} $f(-1)= \i f(1)$ which can be checked directly
by solving the equations of motion at $\varphi=\pi$.
Moreover, it can be seen from \rf{EQstrashny} that
$\i f(-z)$ is its solution if $f(z)$ is a solution.
So we believe
there should be some kind of a crossover where the real solution
becomes
purely imaginary. The fact that the correlation function of a real
quantity becomes imaginary should not disturb us. It means that the
correlation function we are computing may not exist for a certain
value of $\lambda_i$ and can only be understood in the sense of
analytic continuation. Apparently, the logarithm of \PDF which is
obtained by  a Legendre transform of the action we found must remain
real.  One could perform this transform term by term in our expansion.

Summarizing everything, the answer for the D-dimensional case is given by
\be
\left\langle \exp \left\{ \lambda_i \left[ u_i\left({\rho \over 2}\right)
 -
u_i \left( -{\rho \over 2} \right) \right] \right\} \right\rangle =
\exp \left\{
\sqrt
{ \kappa_0 \over 2} {\left( \rho \lambda \right) }^{3 \over 2}
 f \left( \cos \varphi \right) \right\}
\ee
while a Legendre transform of the action will give us the \PDF in
the form
\be
\left\langle \delta \left\{ u_i - \left[ u_i\left({\rho \over 2}\right)
 -
u_i \left( -{\rho \over 2} \right) \right] \right\}  \right\rangle
\approx
\exp \left\{ -  {2 \over 9 \kappa_0}
{u^3 \over \rho^3} \left( 1 + { 9 ( \sqrt{33} - 1) \over {(9 -
\sqrt{33})}^2 }  \ \psi^2 + \dots \right) \right\}
\ee
where $\psi$ is the angle between $u_i$ and $\rho_i$, $\psi \ll 1$,
and $u/\rho \rightarrow
+\infty$.

In conclusion, we would like to say that we have showed by  a simple
computation that WKB calculations are very useful to understand the
behavior of randomly driven Burgers equation and we hope they will
be found useful in other problems of turbulence as well.  The instanton
we found has a spacial homogeneous strain (the velocity was a linear
function in the inertial interval) and we suspect it to be a general
feature of the instantons in the turbulence problem.

We would like to thank A. Polyakov for  his valuable advice and
useful discussions and V. Borue, A. Chekhlov, G. Falkovich, I. Kolokolov,
V. Lebedev and V.
Yakhot for important discussions. One of us (V.G.) is also grateful to
J. Maldacena, R. Gopakumar, M. Moriconi and C. Nayak for useful
conversations.


\begin{thebibliography} {99}
\bibitem{Loop} A. Migdal, {\sl Int. J. Mod. Phys.} {\bf A9} (1994) 1197
\bibitem{Pol} A. Polyakov,  {\sl Phys. Rev. E} {\bf 52}(6)  6183
(1995), hep-th/9506189
\bibitem{FLM} G. Falkovich, I. Kolokolov,
V. Lebedev, and A. Migdal, to be submitted to {\sl Phys. Rev. E}
\bibitem{Jus} J. Zinn-Justin, Field Theory and Critical Phenomena,
(Clarendon Press Oxford, 1989)
\bibitem{Wyl} H. W. Wyld, {\sl Ann. Phys. } {\bf 14} (1961) 143
\bibitem{Yakh} A. Chekhlov, V. Yakhot, {\sl Phys. Rev. E} {\bf 51},
R2739--R2742 (1995); {\sl Phys. Rev. E} {\bf 52}, 5681--5684
(1995)
\end{thebibliography}
\end{document}